# Feedback-assisted Femtosecond Pulsed Laser Ablation of Non-Planar Metal Surfaces: Fabrication of Optical Apertures on Tapered Fibers for Optical Neural Interfaces

Antonio Balena[1,2,*], Marco Bianco[1,2], Filippo Pisano[1], Marco Pisanello[1], Leonardo Sileo[1], Bernardo Sabatini[3], Massimo De Vittorio[1,2,+], Ferruccio Pisanello[1,+]

[1] Fondazione Istituto Italiano di Tecnologia, Center for Biomolecular Nanotechnologies, Arnesano (LE), 73100, Italy
[2] Dipartimento di Ingegneria Dell'Innovazione, Università del Salento, Lecce, 73100, Italy
[3] Department of Neurobiology, Howard Hughes Medical Institute, Harvard Medical School, Boston, MA 0211.
+ Equally Contributing
*antonio.balena@iit.it

**Abstract:** We propose a feedback-assisted direct laser writing method to perform laser ablation of fiber optics devices in which their light-collection signal is used to optimize their properties. A femtosecond-pulsed laser beam is used to ablate a metal coating deposited around a tapered optical fiber, employed to show the suitability of the approach to pattern devices with small radius of curvature. During processing, the same pulses generate two-photon fluorescence in the surrounding environment and the signal is monitored to identify different patterning regimes over time through spectral analysis. The employed fs beam mostly interacts with the metal coating, leaving almost intact the underlying silica and enabling fluorescence to couple with a specific subset of guided modes, as verified by far-field analysis. Although the method is described here for tapered optical fibers used to obtain efficient light collection in the field of optical neural interfaces, it can be easily extended to other waveguides-based devices and represents a general approach to support the implementation closed-loop laser ablation system of fiber optics.

## 1. Introduction

Ultrafast lasers opened new paradigms to material processing thanks to the high peak intensity achievable with focused pulses at relative low energies[1]. Femtosecond laser ablation and micromachining has been applied both on dielectric and on metallic samples[2–4], and it has been used for the realization of laser-inscribed metallic surface structures such as micro-holes[4], micro-grooves[5], and micro-pillars[6]. Laser ablation has been also applied to optical fibers to obtain side emission of light by drilling micro discontinuities inside the fiber's core[7,8]. Although this process enables multi-site emission of light, it compromises the light propagation and collection properties of the waveguide itself, hence affecting its performances in sensing applications. This is a particularly relevant factor in applications where modal properties of the waveguide should be preserved while maximizing the light collection efficiency of the fiber.

Sensors based on Tapered Fibers (TFs) are an example[9] that has found a range of applications over the years, including sensing of biomolecules[10,11], and, very recently, implantable systems to optically monitor neural activity[12]. In this latter case, the taper does not shrink at its central portion, as in canonical tapered fibers sensors based on transmission measurements, but it interrupts at its narrower section, creating a tip that enables for a smooth implantation in the brain and functional fluorescence collection[12]. The tip is then coated with reflective metal and a set of optical windows are opened along the narrowing waveguide, enabling guided modes of different order to interact with different regions of the brain. This approach enables micro-structured TFs (µTFs) to deliver and collect light with depth resolution along a single probe, a unique feature for optical neural interfaces[12–16]. Accordingly, tailoring the shape and size of the optical windows enables to engineer light delivery and collection volumes in order to collect functional fluorescence signal from groups of few neural cells[12]. However, it is important that the structuring of the narrowing region preserves the effect of the waveguide on propagating modes, not to compromise the optical interaction between the environment and the modal content[17].

The fabrication of µTFs with optical windows can be carried out with high-resolution fabrication techniques including Focused Ion Beam (FIB) Milling[18,19], as well as fast-writing methods such as Ultraviolet Direct Laser Writing (UV-DLW)[20]. This latter approach allows for high-speed patterning of the metal coating at the cost of limited resolution and higher roughness of the window profile. However, both FIB and UV-DLW unavoidably undermine the underlying dielectric surface. Although this does not compromise the multi-site light delivery capability of the TFs, light collection efficiency would greatly benefit from an intact surface interacting with incoming fluorescence signal.

In this work, we propose a Feedback-Assisted Direct Laser Writing (FA-DLW) method to pattern the edge of a TF to optimize light collection properties for applications in the field of optical neural interfaces. The method takes advantage of laser ablation of the metal layer on the highly non-planar surface of the taper while it is submerged in a fluorescent bath. A scanning Near Infra-Red (NIR) femtosecond pulsed laser ablates the region of interest and simultaneously generates a two-photon fluorescent signal in the bath. Continuously monitoring this signal provides a spectral fingerprint that identifies when the metal is totally removed. The laser pulses interact mostly with the metal coating, leading to its ablation, while leaving the underlying glass surface intact. This enabled the production of optical windows that couple light in defined subset of guided modes, as verified by far-field analysis on the distal facet. The result is a fabrication approach that preserves the waveguides modal properties while optimizing the collection efficiency of the device.

## 2. Feedback-assisted fabrication of micro-structured TFs

### 2.1 FA-DLW Setup

The particular advantage of the proposed method is the ability to monitor the ablation progress in real-time. In order to record the collection properties of the optical apertures during the fabrication, we designed a Two-Photon (2P) ablation system assisted with a feedback signal (Fig. 1A). A fs pulsed NIR Laser (Coherent Chameleon Discovery, average output power at 800 nm of 1.8 W) was modulated by a Pockels Cell. A quarter wavelength plate was used to obtain circular polarization, and the beam was expanded to fit the size of a 6 mm galvanometric mirrors pair (GMs) employed to deflect the beam over the (x,y) plane. The beam was then expanded and collimated by a scan-tube lens assembly and sent on the back aperture of a 4x 0.28 NA objective lens (OBJ1). The setup was controlled by a DAQ system interfaced with the open source software Vidrio ScanImage.

The laser spot was raster scanned in the field of view of OBJ1 (2085x2085 $\mu m^2$) (Fig.1.A) generating a fluorescence spot into a fluorescent bath (30μM $C_{20}H_{12}O_5$ (Fluorescein) in Phosphate buffered saline (PBS)), in which the aluminum-coated tapered fiber was submerged. The fluorescence collected by OBJ1 was reflected by a dichroic mirror (D1) and detected by a Photo-Multiplier Tube (μscope-PMT). We used the resulting 2P fluorescence image to position the beam over the area to be ablated.

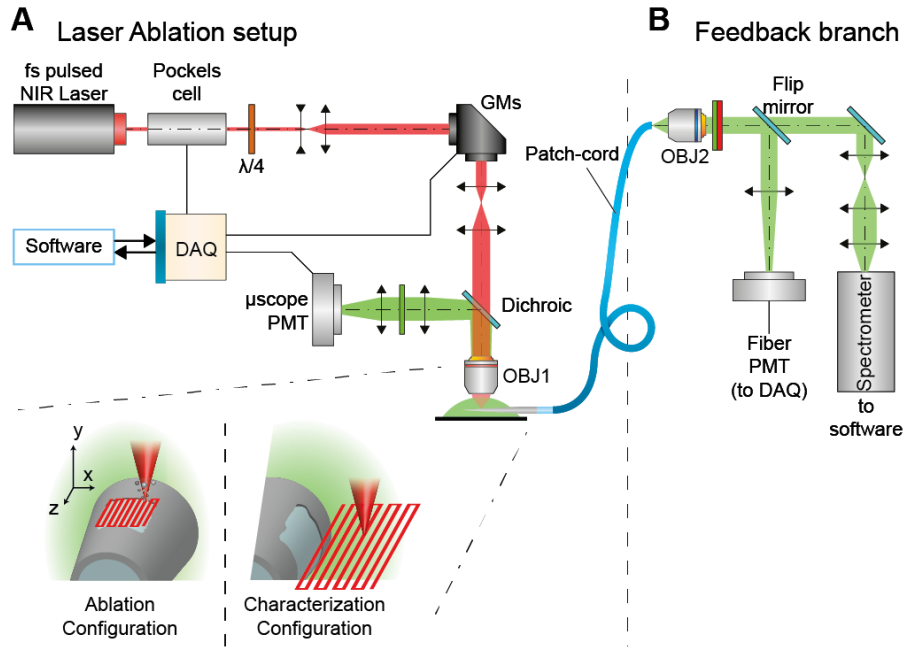

Fig. 1. (A) Custom Two-Photon Microscope for spatially restricted Laser Ablation of Aluminum deposited on TFs and for the optical characterization of the resulting devices. The zooms show sketched details on the raster scanning trajectories of the Laser spot in two different configurations: the Ablation Configuration on the left and the Characterization Configuration on the right. (B) Fluorescence Collection optical setup that allows for real-time Feedback of the Ablation process and the related optical characterization.

A second collection path through the TF was employed to monitor ablation, exploiting light entering the waveguide across the ablated pattern (referred to as Feedback Branch in Fig.1B). Light guided by the taper propagated in the patch-cord and entered the feedback branch by an objective lens (OBJ2). The signal was filtered by a band-pass (central wavelength 525 nm, bandwidth 39 nm) and a NIR block filters (cutoff wavelength 770nm, short pass), and was directed toward two different optical detectors. A Spectrometer offered a real time feedback during fabrication, while a second PMT (Fiber-PMT) allowed for measuring a collection intensity map from the window just after the ablation.

## 2.2 Optical Windows Fabrication

TFs were obtained from OptogeniX (www.optogenix.com), and were produced from NA=0.39 fibers (Thorlabs FT200UMT) with core/cladding diameter of 200µm/225µm through a heat and pull procedure, described in detail by Sileo et al.[19]. Thermal evaporation was then employed to conformally deposit a 400 nm-thick Aluminum layer around the taper, setting the deposition rate at about 1.5 Å/sec while the TF was continuously rotated with a stepper motor.

The size of the windows was set by restricting the scanning region of the GMs to a 40x40 µm$^2$ area in correspondence of a taper diameter of 60 µm. Fabrication parameters were chosen considering two factors. (i) The laser fluence has to be higher than the ~100 fs pulse ablation threshold of Aluminum, found to be $F_{t,Al}$ ~ 0.1 J/cm², [21,22] but lower than both the damage threshold and the ablation threshold for fused silica, found to be, respectively, $F_{damage,SiO2}$ > 2 J/cm² and $F_{t,SiO2}$ > 3 J/cm².[23] Hence, we set the laser fluence at F ~ 0.2 J/cm². Working close to the threshold enables an ablation rate of ~2-6 nm/shot in air[22], with the liquid environment further reducing this rate.[24] (ii) A wavelength which could efficiently excite 2P fluorescence in the feedback fluorescent solution had to be chosen, hence we set λ = 800 nm. We configured a dwell time on each raster pattern's point of $t_{dwell}$=3.2 µs (about one frame per second with 512×512 pixels).

Representative spectra recorded during the process are reported in Fig. 2A-B as a function of time. Although the processing of a 40x40 µm² optical window can be obtained in about 90 seconds, the spectra reported in Fig.2A-B were recorded with a laser fluence near to $F_{t,Al}$, in order to slow down the fabrication and to better display the evolution of the process through the collected fluorescence signal. From the beginning of the fabrication, the spectra grow in intensity from an initial near-to-zero flat shape to eventually unfold into the typical fluorescein emission spectrum,[25] with no additional spectral features clearly recognizable. The integrated intensity values versus time, displayed in Fig. 2C, show three different regimes (defined here as 'regime I' for 0 < t < 75 s, 'regime II' for 75 s < t < 750 s, and 'regime III' for t > 750 s).

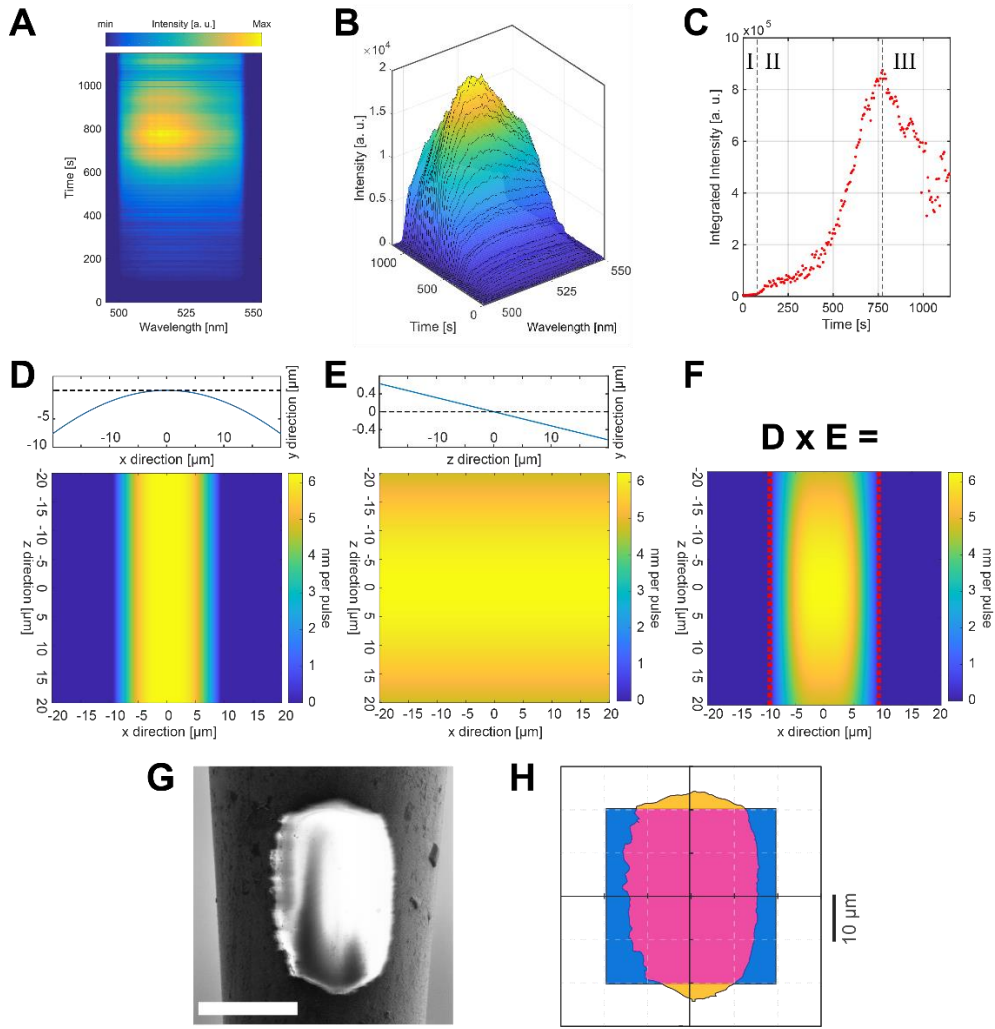

Fig. 2 (A) Real-time time-resolved fluorescence collection map used as feedback of the ablation process. (B) 3D waterfall plot of the collection spectra as a funcion of time from which the spectral shape of the collected signal is observable. (C) Scatter plot of the spectra integrated intensity versus time, with the beginning of the process being at t = 0). The dashed lines divide the plot into three different regimes, described in the main text. (D) Ablation depth per pulse dependence from the TF curvature, determined through Eq. 1. (E) Ablation depth per pulse dependence from the axial inclination of the TF, determined throug Eq. 1. (F) Resulting effect on the ablation depth per pulse. Red dashed lines enclose the non-zero ablation depth per pulse region. (G) Scanning Electron Micrograph of a FA-DLW window at a diameter of ~60 μm. Scale bar is 25 μm. (H) Difference between the Field of View and the final FA-DLW window surface.

During 'regime I', no light is coupled inside the waveguide because the aluminum layer has not been sufficiently ablated yet. This is followed by 'regime II' in which the integrated intensity progressively increases. This corresponds to progressing metal ablation from the region of interest. During 'regime III' the integrated intensity starts decreasing and oscillating, although a constant fluorescence signal could have been expected after the complete ablation of the metal. We attributed this behavior to the generation of cavitation bubbles in the liquid environment near the focal spot, due to local heating[26–28]. Since the laser is focused on the fiber surface, the bubbles grow in the vicinity of the collection region, hence reducing the fluorescence signal that can be collected by the feedback system.

Since the ablation takes place on a curved surface while the laser scans in a plane, the local effective fluence depends on the cosine of the incidence angle[29]. Thus, the radius of curvature of the taper imposes a variation of the ablation depth per pulse as a function of the surface profile versus the planar scan plane, following [29]:

$$\Delta y(x) = \alpha^{-1} \log\left\{\frac{F}{F_t}\cos\left[\tan^{-1}\left(\frac{dy}{dx}\right)\right]\right\} \qquad \text{[Eq. (1)]}$$

Where $\alpha = 4\pi\kappa/\lambda$ is the effective absorption index of Aluminum, and $\kappa = 7.05$ is its extinction index at $\lambda = 800$ nm taken from Ref.[30].

We calculated these variations considering the TF profile in the x-direction (Fig. 2D) and the taper angle along the z axis (Fig. 2E). Ablation could take place only for positions at which the ablation depth per pulse is positive (Fig. 2F). From these calculations we expect that the window size is minimally affected by variations in the z-direction (expected z-size 40 μm) but that it changes considerably along the x-direction (expected x-size ~20 μm). A representative Scanning Electron Micrograph of a FA-DLW window is shown in Fig. 2G. From the SEM image it is clear that the shape of the window is similar to the expected one outlined by the iso-depth lines in Fig.2F, with the final size differing by about 5 μm in each direction. This is because Eq. 1 does not consider the actual size of the laser spot, especially along the y direction, while our system has a PSF with lateral FWHM of 3 μm and axial FWHM of ~30 μm[31]. Moreover, residual thermal energy deposition could affect the ablated region, enlarging its edges[32,33].

The software MATLAB Image Segmentation tool was employed to calculate the effective area of the window and to compare it with the size of the ablation Field of View (Fig. 2H). The actual size was estimated in 1375 μm$^2$, resulting in the 85% of the target 1600 μm$^2$.

## 3. Collection properties characterization

The possibility to collect light with μTF is enabled by the dielectric openings, e.g. the optical windows, on the surface. Let us consider a dielectric TF (Fig. 3A) and a description of propagating energy based on linearly polarized (LP) modes. The transversal propagation constant of the *l,m*-th guided mode at the section of radius *r(z)* is given by the following relation [34]:

$$k_{t-l,m}(z) \simeq \frac{r(z=0)}{r(z)} k_{t-l,m}(z=0) \qquad \text{[Eq. (2)]}$$

where $k_{t-l,m}(z=0)$ is the transversal propagation constant of the *l,m*-th guided mode at the widest taper section. In this dielectric confinement regime, the *l,m*-th mode is propagative only if its transversal propagation constant fulfill the condition $k_{t-l,m}(z) < k_0 NA$. Therefore, since smaller *r(z)* implies higher $k_{t-l,m}(z)$, a cutoff radius for the *l,m*-th mode can be defined ($r(z_{c-l,m})|_{dielectric}$), and for $r(z)<r(z_{c-l,m})|_{dielectric}$ the mode can couple with light in the environment (Fig. 3A).

In metal coated TFs equation (1) holds true, but the *l,m*-th mode is guided under the evanescence condition $k_{t-l,m} < k_0 n$, where *n* is the refractive index of the waveguide. This sets a different cutoff section for metal-coated TFs $r(z_{c-l,m})|_{metal} < r(z_{c-l,m})|_{dielectric}$ since *n>NA* (Fig. 3B)[15,35]. As a consequence, light entering an optical window realized at a section with radius *r* along the taper will couple with guided modes having a $k_{t-l,m}(z=0)$ such that they can propagate in a metallic waveguide.

Because of this mechanism, two different characterizations are required to fully describe the light collection properties of the window: (i) to estimate the efficiency of photons collection through the aperture and (ii) to determine how photons are coupled to guided modes. Thus, we characterized the near-field properties in terms of collection efficiency maps (η)[12,31], as well as the far-field pattern of light emerging from the distal facet, obtaining direct access to the transversal propagation constant of modal content back-propagating into the fiber[12,15]. The two approaches are described in following paragraphs 3.1 and 3.2, respectively.

### 3.1 Collection efficiency maps

Collection efficiency maps (η(x,y)) represent the percentage of photons collected by the window when an omnidirectional point source emits in the generic point of the space (x,y). To obtain a direct measurement of η for the FA-DLW windows (Fig. 3C), we placed the fiber inside a drop of PBS:Fluorescein solution making sure that the normal to the window's surface lied perpendicularly to the optical axis of the Objective Lens (Characterization Configuration in Fig. 1A).

A λ = 920 nm fs-pulsed NIR laser beam was focused and scanned in a ~ 260 x 260 µm² area, in order to obtain the scan of a point-like fluorescence source in close proximity to the window.

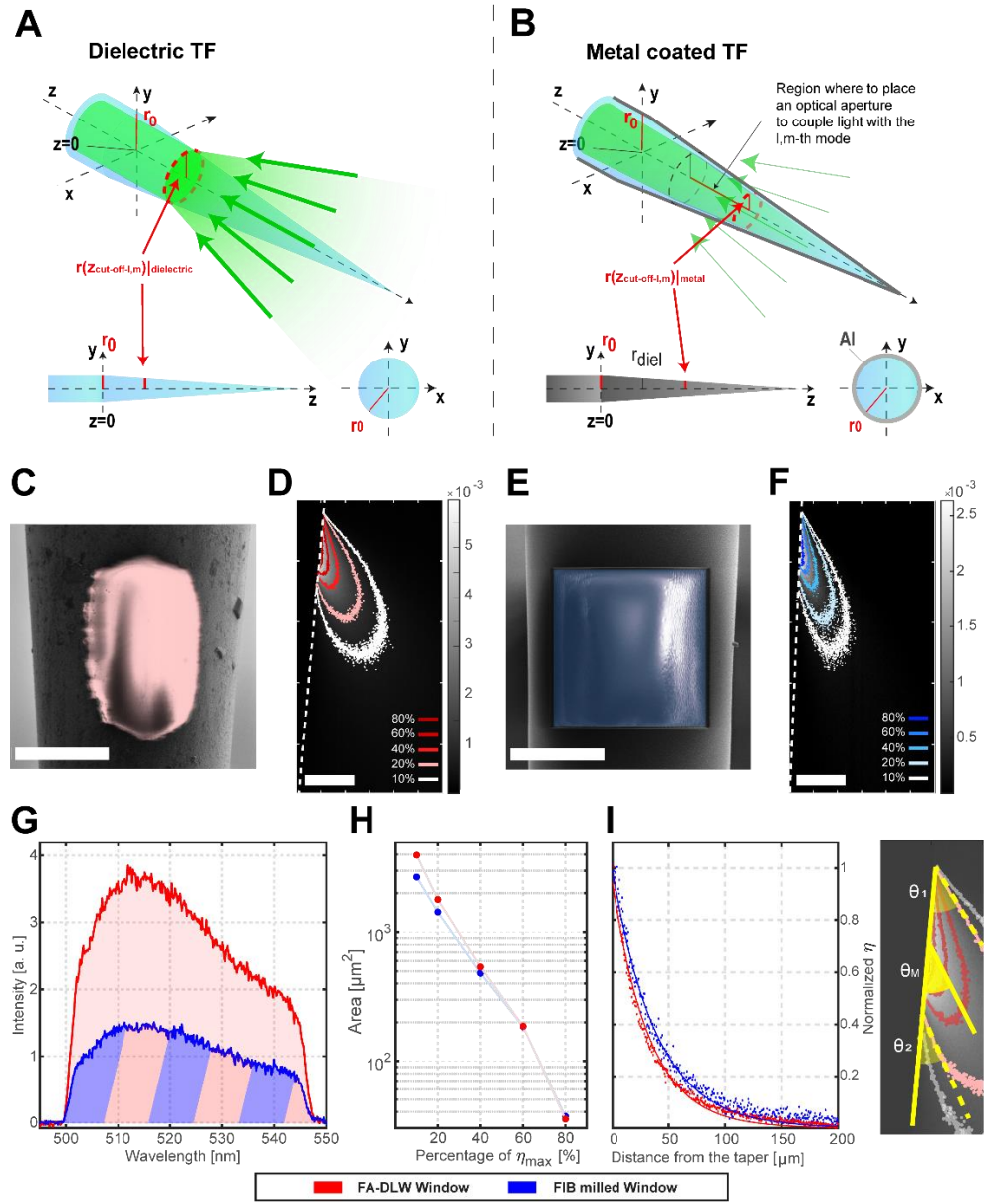

Fig. 3 (A) Schematic representation of a dielectric TF. (B) Schematic representation of a metal coated TF. (C) Scanning Electron Micrograph of a FA-DLW window. Scale bar is 25 μm. (D) Collection efficiency map for FADLW window (red). Scale bar is 20 µm. (E) Scanning Electron Micrograph of a 40x40 µm$^2$ FIB milled window. Scale bar is 25 μm. (F) Collection efficiency map for FIB milled window (blue). Scale bar is 20 μm (G) Spectra registered for the two type of windows. (H) Collection areas at different percentages of the maximum collection efficiency value. (I) Decay profile for normalized collection efficiencies. The decay is measured along a straight line which starts from the center of the window and forms with the taper edge (white dashed lines in Fig 3.D,F) an angle equal to $\theta_M$, that is the mean between the angles formed by the taper edge and the two edges of the 20% iso-surface profile ($\theta_1$ and $\theta_2$, example on a detail from Fig. 3.D).

The fluorescence signal collected by the window was recorder by the Fiber-PMT in terms of number of photons per pixel ($N_f(x,y)$), while the µscope-PMT builds an image of the same area, also in this case in terms of number of photons per pixel $(N_s(x,y))$[31]. Considering the detection loss of the system, η was therefore determined by the following relation:

$$\eta(x,y) = \frac{N_f(x,y)}{Q \cdot N_s(x,y)}$$ [Eq. (3)]

Where Q is a factor that takes into account the detection loss of the epifluorescence collection path and the term $Q \cdot N_s(x,y)$ represents the number of photons emitted by the excited fluorescence spot.

A representative collection efficiency map is displayed in Fig. 3D, showing iso-intensity lines in the x,y plane for different values of η. The maximum η achieved by the single window was estimated by summing all the pixel values with η > 90% of the maximum pixel value and dividing for the corresponding number of pixels, obtaining $\eta_{max} = (5.50 \pm 0.13) \times 10^{-3}$ (mean ± standard deviation, n=3 fibers). We carried out a quantitative comparison between the FA-DLW window and a Focused Ion Beam milled window realized at the same diameter on a different metal coated TF (Fig. 3E)[18], milling an area equal to the ablation FOV (1600 μm$^2$). This resulted in the η maps displayed in Fig. 3F and a $\eta_{max} = (2.30 \pm 0.06) \times 10^{-3}$ (mean ± std, n=3 fibers). The integrated intensity detected by FA-DLW spectrum (red) resulted to be ~2.6 times higher than the FIB spectrum (blue), as confirmed by the spectra in Fig. 3G.

To better compare the properties of the FA-DLW and FIB windows, we measured the surface area defined by the isolines at different percentages of $\eta_{max}$ (Fig. 3H). While the high relative efficiency areas (60% and 80% of $\eta_{max}$) are similar, the low relative efficiency area (10% of $\eta_{max}$) is 1.5 times wider for the FA-DLW one (3956 μm$^2$ versus 2676 μm$^2$). If the decay profiles of the normalized η are considered (Fig. 3G), two similar trends can be observed, with the efficiency that drops below the 20% of the maximum after the first 50 μm after the window.

This lets us suggest that FA-DLW windows are better suited to detect optical signals than FIB-milled windows, as the higher collection efficiency of the formers allows gathering a stronger signal from larger sample volumes.

## 3.2 Angle Selective Light Coupling setup and Far-Field Imaging

To identify the transversal propagation constant of modal content excited by light coupled through the window, we have implemented the optical path in Fig. 4 A, B. The patterned TF was placed in a PBS:Fluorescein bath, and a Continuous Wave laser at 473 nm was injected into the fiber with a specific angle to maximize output power from the window. Generated fluorescence was collected by the same window and it was directed through a far-field imaging path by a dichroic mirror.

Light emitted by the fiber facet can be expressed as a weighted sum of plane wave components $w(x,y)$ which propagate at manifold $(\theta_{xz}, \theta_{yz})$ angles (in the scheme in Fig. 4.C, only one plane wave is shown). Passing through lens $L_3$, those components are separated and focused at different points $\mathbf{r}(u,v) = (f_3 \cdot \tan(\theta_{xz})\mathbf{u}, f_3 \cdot \tan(\theta_{yz})\mathbf{v})$ on the focal plane, hence being directly correlated with the angular distribution of light, which is intrinsically related to the transversal wave-vector $k_t$ emitted from the fiber[12,36]

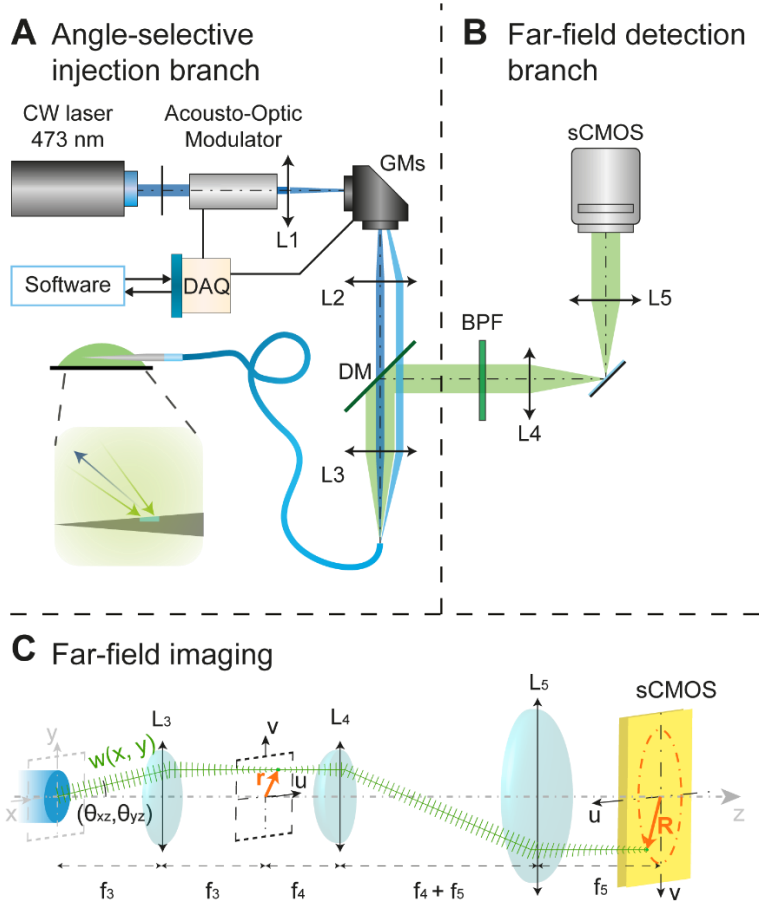

Fig. 4 (A) Sketch of the light emitted by the fiber and the correlation between the emission angle and the coordinates on the focal plane. (A) Scheme of the optical setup employed for angle selective light coupling inside the fiber and (B) far-field imaging of the collected signal. (C) Sketch of the light emitted by the fiber and the correlation between the emission angle and the coordinates on the detection plane (Band Pass Filter BPF is omitted).

An additional relay with lenses L4 and L5 matches the size of the far-field image in the $(u,v)$ plane with the size of the imaging sensor, with the image being related to $k_t$ values by Eq. (4):

$$k_t = \frac{2\pi}{\lambda} \sin\left[ \tan^{-1}\left( \frac{f_4}{f_3 f_5} |\mathbf{R}(u,v)| \right) \right] \qquad \text{[Eq. (4)]}$$

where $\mathbf{R}(u, v)$ is the magnitude of the distance vector of the pixel from the center of the resized (u, v) plane, f3, f4 and f5 are respectively the focal lengths of L3, L4 and L5 in Fig.3.B. Because of the cylindrical symmetry of the modes' propagation into the waveguide, we observed a ring-shaped intensity distribution pattern on the detection sensor.

Representative far-field patterns of light detected from a window placed at a diameter d = 55 μm is displayed in Fig. 5A, after background subtraction to avoid the influence of patch fiber auto-fluorescence on the measurement (the background image was obtained in a non-fluorescent PBS bath). The image was numerically segmented by: (i) a low-pass intensity threshold to remove the residual excitation laser and outliers, and (ii) a high-pass intensity filter to remove residual noise (placed at two times the mean value of the background pixels).

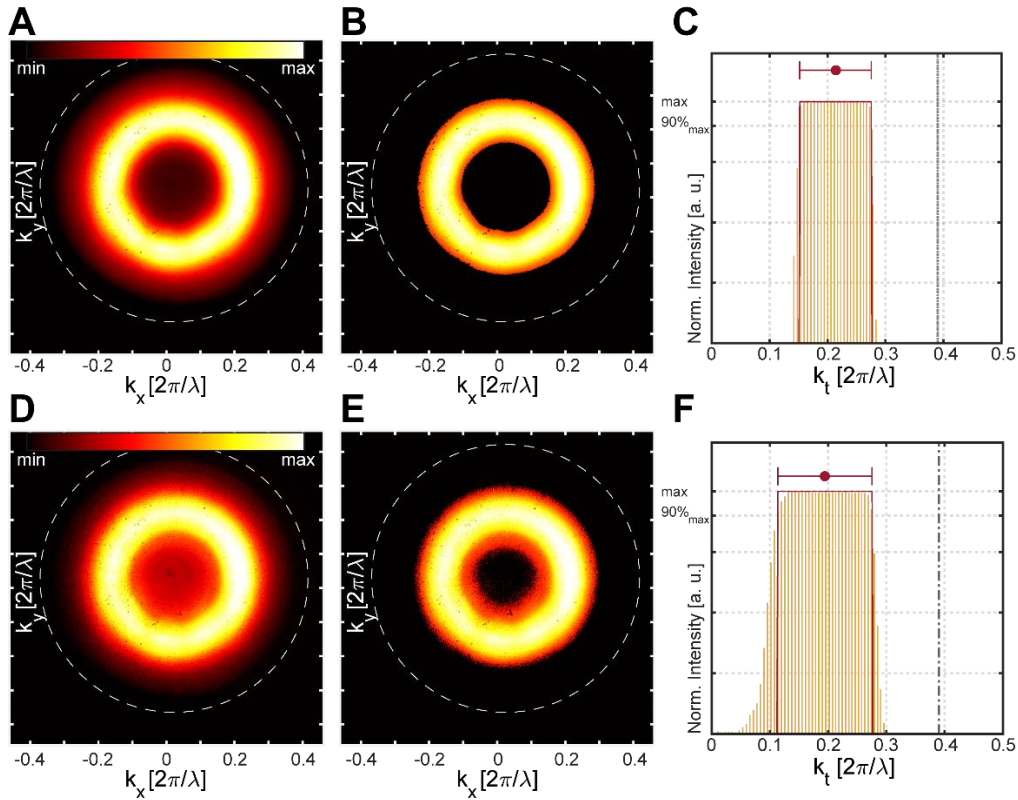

Fig. 5 (A) "Raw" Far-Field pattern obtained from the subtraction between the signal collected while the fiber is submerged in a PBS:Fluorescein solution drop and the signal collected while the TF is submerged in non-fluorescent PBS. (B) Ring-shaped pattern obtained from the algorithm. (C) (Yellow) Histogram of the distance in units of $2\pi/\lambda$ between the non-zero pixels of the image corrected as described in the main text and the centroid of the image (only a bar each 4 is shown for visibility). (Red) Gaussian fit of the histogram. The dot-dashed line corresponds to the maximum $k_t$ that could be emitted by the fiber (0.39 $2\pi/\lambda$). (D-F) As in A-C

for a FIB milled window. White dashed circles the maximum $k_t$ that could be emitted by the fiber (0.39 $2\pi/\lambda$).

A representative segmented image is shown in Fig. 5B. Subsequently we plot the histogram of the $k_t$ values related to non-zero pixels from the centroid of the segmented image (Fig. 5C). The $k_t$ interval of the modal subpopulation outcoupled from the patch cord is then extracted from the histogram as the values greater than the 90% of the maximum (red dot in Fig. 5C). From this analysis we obtained $k_t = 0.21 \pm 0.06$ [$2\pi/\lambda$] for FA-DLW windows. We repeated the experiment for the FIB milled window (Fig. 5D, 5E), and we extracted the outcoupled $k_t$ distribution from the histogram (Fig. 5F), obtaining $k_t = 0.19 \pm 0.08$ [$2\pi/\lambda$].

As expected, the collection diameter determines the modal subset at which the collected light propagates in the fiber. From the comparison of far field images, we observed that FIB-milled and FA-DLW milled windows collect modal subsets with compatible $k_t$. Therefore, the FA-DLW window shows higher collection efficiency, while preserving unchanged the modal coupling properties of the waveguide, an important feature for depth-selective fluorescence collection[12].

## 4. Conclusions

We propose an ultrafast Direct Laser Writing setup that takes advantage of real-time spectral feedback to optimize light collection properties of microstructures realized on the edge of a fiber optic-based device. The system was applied to ablate a conformal metal coating on the edge of a tapered fiber, while monitoring the two-photon fluorescence signal simultaneously generated in the environment by the same fs laser pulses. When compared with canonical FIB-milled patterns, a quantitative analysis of the optical properties of collected light in both near-field (collection efficiency maps) and in far-field (outcoupled modes dispersion analysis) show ~60% improvement in the collected signal, while the uncoupled modal subset is comparable between the fabrication processes.

We attributed these findings to the lower alteration of glass induced from the laser ablation process, in terms of both roughness and dielectric properties. From the SEM micrographs (Fig. 3C, E), we observe no notable texture in the active area of the FA-DLW window. On the other hand, FIB processing shows no textures in the central part, for an extension of ~20 μm in the transversal direction with respect to the TF axis, but a rough surface in the lateral part, where the curvature of the fiber is higher. In addition, the FIB milling on glass is known to induce a variation of its optical properties, with a refractive index rise from *n=1.46* to *n~1.75* in the visible range, and therefore a transmissivity decreases from *T=90%* to *T=80-85%*, due to Ga$^+$ ions implantation[37]. Moreover, the feedback offered by the developed system, as well as the possibility to characterize optical apertures' properties using a single setup, sensibly reduce the overall fabrication time with respect to FIB.

As µTFs have proven their potential in depth-resolved photometry experiments[12], we envision that the presented FA-DLW fabrication of optical apertures on the edge of the TF surface could be employed to engineer the collection sites along the TF itself. Indeed, spatial-selective photometry in high-scattering media, such as the brain, could benefit from this type of processing, as it requires high collection efficiency. Although here we showed FA-DLW in the field of optical neural interfaces, the same concepts can be extended to a range of fiber-based biosensors, which would greatly benefit of optimized and selective interactions of guided modes with the environment.

**Funding**

A.B., M.B., F. Pisano and F. Pisanello acknowledge funding from the European Research Council under the European Union's Horizon 2020 research and innovation program (G.A.677683). M.P. and M.D.V. acknowledge funding from the European Research Council under the European Union's Horizon 2020 research and innovation program (G.A. 692943). F. Pisanello and M.D.V. acknowledge funding from the European Union's Horizon 2020 research and innovation program under grant agreement No 828972. M.D.V. is funded by the US National Institutes of Health (U01NS094190). M.P., L.S., F. Pisanello and M.D.V. are funded by the US National Institutes of Health (1UF1NS108177-01).

**Disclosures**

LS, MDV, BS and F. Pisanello are founders and hold private equity in Optogenix, a company that develops, produces and sells technologies to deliver light into the brain. Tapered fibers commercially available from Optogenix were used as tools in the research.